\documentclass{ws-ijmpa}

\begin{document}

\markboth{U. THOMA}
{PHYSICS AT ELSA, ACHIEVEMENTS AND FUTURE}

%
\catchline{}{}{}{}{}
%

\title{PHYSICS AT ELSA, ACHIEVEMENTS AND FUTURE}

\author{\footnotesize U.THOMA}

\address{2. Physikalisches Institut 
der Universit\"at Giessen, \\
Heinrich-Buff-Ring 16, 35392 Giessen and \\
Helmholtz-Institut f\"ur Strahlen- und Kernphysik der \\Universit\"at
Bonn, Nu{\ss}allee 14-16, 53115 Bonn, Germany
}

\maketitle


\begin{abstract}
At ELSA interesting results  on baryon resonances have been obtained 
by the CB-ELSA, the CBELSA/TAPS and the SAPHIR collaborations. 
New resonances were found, in particular a new $\rm D_{15}(2070)$ decaying into
$p\eta$, was recently observed by the CB-ELSA experiment. 
The availability of a polarized beam and a 
polarized target did allow to measure the GDH sum rule up to 2.9 GeV. 
In the future
double polarization experiments will be
performed using the Crystal Barrel detector together with new forward
detector components. These polarization observables will provide important
additional information for the partial wave analyses performed to extract the
contributing resonances and their parameters from the data.  

\keywords{Baryon spectroscopy, missing resonances, GDH sum rule}
\end{abstract}

\section{Experiments at ELSA}	
The electron stretcher ring ELSA in Bonn accelerates  
electrons up to a maximum energy of 3.5~GeV. 
These have been used for photoproduction experiments with either
unpolarized or linearly polarized photons, the latter being
produced via coherent bremsstrahlung. For the GDH-experiment 
polarized electrons with an energy up to 3~GeV were used together with
a polarized target to measure the helicity dependent total photoabsorption cross
section.  
The GDH-detector is
optimized for inclusive measurements. Its 
overall acceptance for hadronic processes is better than 
99$\%$ . The electromagnetic background is suppressed by about five orders of
magnitude by means of a threshold Cherenkov detector\cite{gdh_helbing}. 
The CB-ELSA and the CBELSA/TAPS
collaborations both used the Crystal Barrel calorimeter as central
detector. In the CB-ELSA setup it did consists of 1380 CsI(Tl)
crystals covering 98$\%$ of the 4$\pi$ solid angle. In the CBELSA/TAPS
setup the  Crystal Barrel calorimeter was opened up in forward
direction to $\pm 30^{\circ}$. This forward angular range was
covered with the TAPS detector consisting out of 528 $\rm BaF_2$
crystals with plastic scintillators in front. Both setups are very
well suited to measure photons and reach an almost 4$\pi$ angular
coverage. 
In contrast to this the SAPHIR detector is well suited to
detect charged particles.
The CB-ELSA, the CBELSA/TAPS and the SAPHIR detector setups have 
been used to measure exclusive hadronic channels.
In the following some of the results
obtained by the  different experimental setups will be discussed. 
\section{Measurement of the GDH sum rule}
The Gerasimov-Drell-Hearn (GDH) sum rule, 
\begin{equation}
\int_0^{\infty}  \frac{d\nu}{\nu} ( \sigma_{3/2}(\nu)-\sigma_{1/2}(\nu)
    ) = \frac{2
    \pi^2 \alpha}{m^2} \kappa^2 \, , 
\label{diseqn}
\end{equation}
which connects the anomalous magnetic moment of the
nucleon - a static property - with the spectrum of its excited states,
has been measured at ELSA and MAMI. 
$\kappa$ and m are the anomalous magnetic moment and the mass of the nucleon,
respectively. 
\begin{figure}[h!]
\vspace*{-1.2cm}
\begin{tabular}{c}
\hspace*{+.1cm}
\epsfig{file=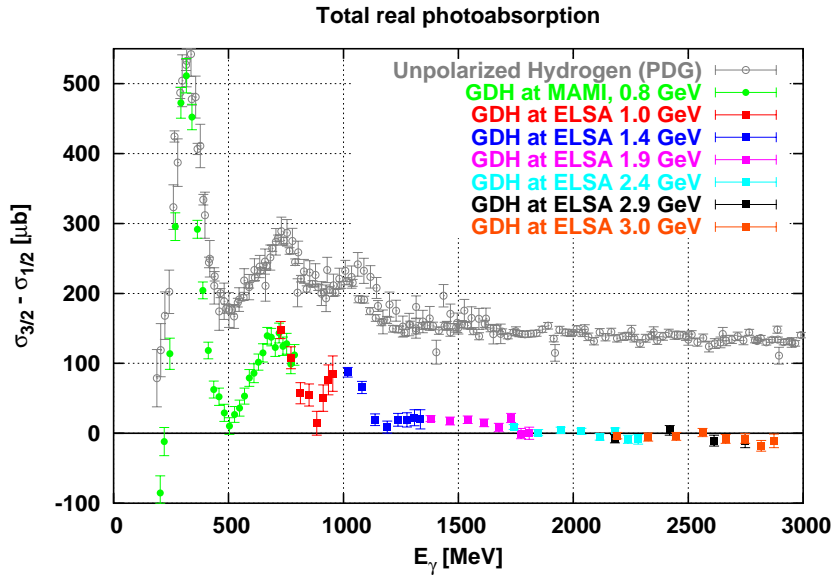,height=0.9\textwidth,angle=-90,clip=} \\[+1ex]
\end{tabular}
\vspace*{-0.4cm}
\caption{Difference of the polarized total photoabsorption of the
  proton as measured at at ELSA and
  MAMI\protect\cite{gdh_proton} in comparison to the unpolarized cross
  section\protect\cite{pdg}. The picture is taken from\protect\cite{gdh_neutron} 
}
\vspace*{-0.2cm}
\label{all_wq}
\end{figure} 
The GDH-integral measured at ELSA and MAMI 
is extrapolated to lower energies by MAID and to higher
energies using either a Regge-parametrisation or a fit to the
GDH-data\cite{gdh_proton}. The value for the GDH integral obtained is within the errors consistent
with the expected value. 
Some first data on the neutron has also been taken\cite{gdh_neutron}.

\section{Baryon spectroscopy}
\begin{figure}[t!]
\begin{tabular}{rl}
{\includegraphics[width=.55\textwidth]{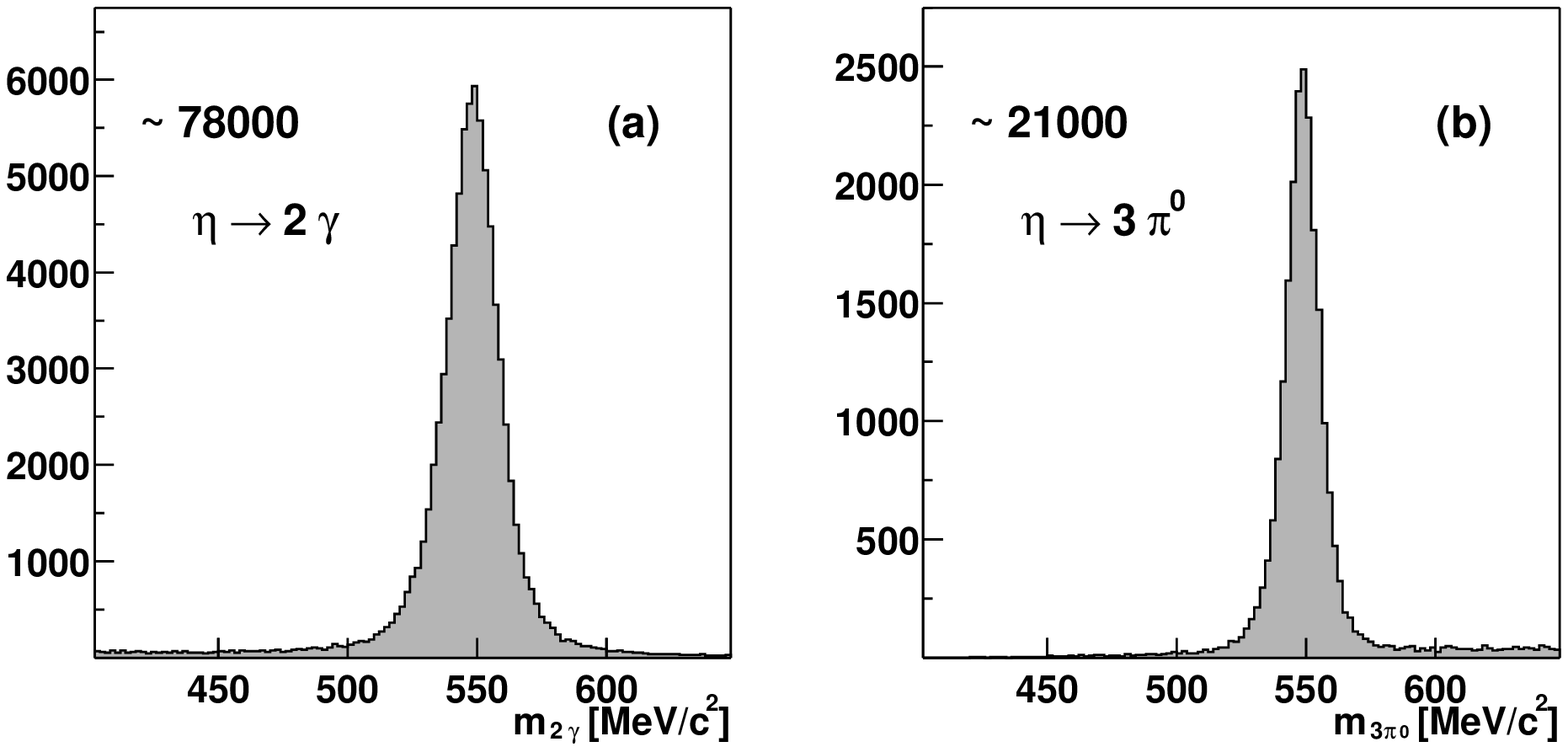}} 
&  \hspace*{-0.8cm} {\includegraphics[width=.50\textwidth]{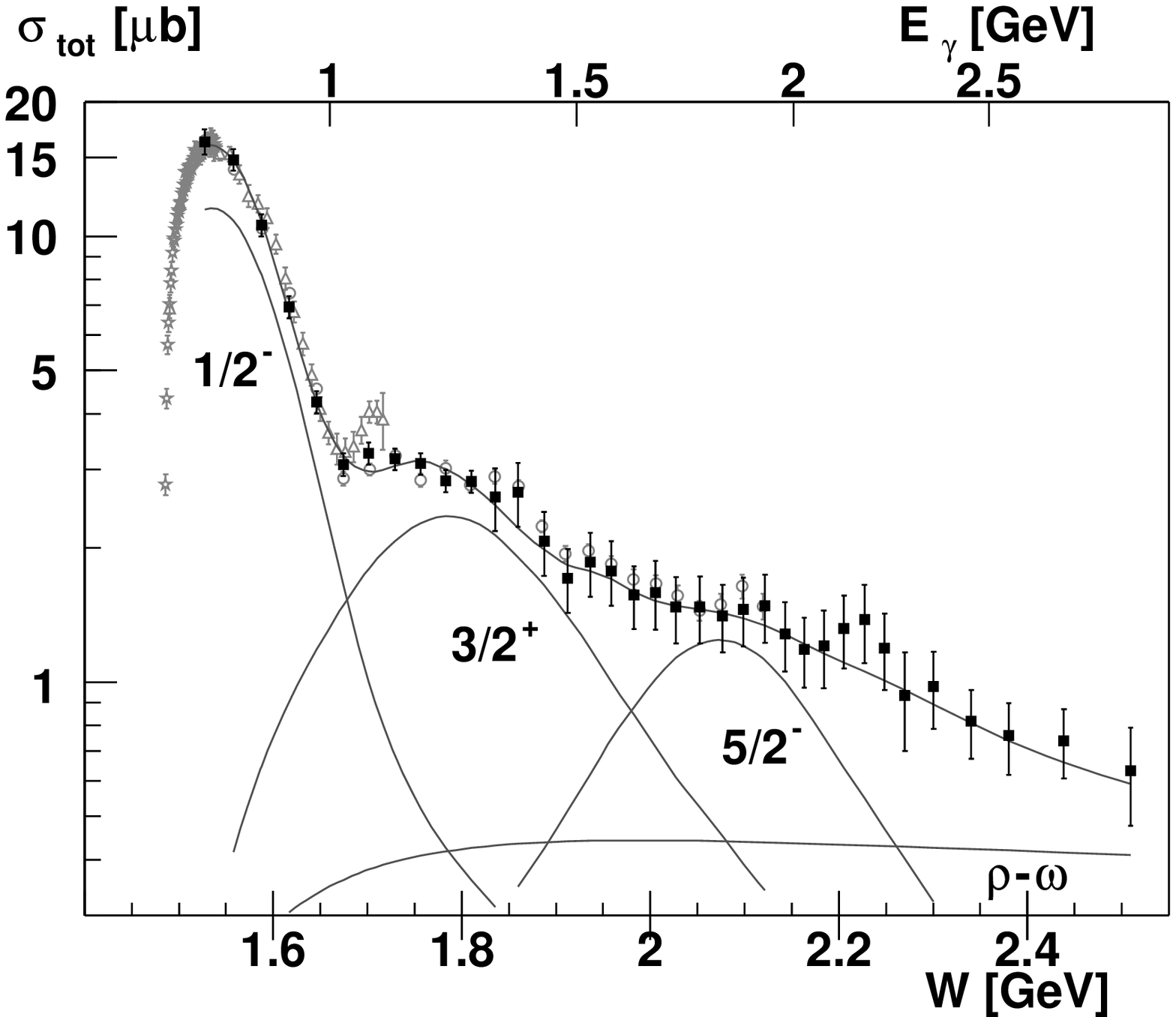}} \\ 
\end{tabular}
\caption{Left two plots: Invariant $\gamma\gamma$ and $3\pi^0$ invariant mass. 
Right: Total cross section (logarithmic scale) for the reaction
$\gamma\,\rm{p}\rightarrow\rm{p}\,\eta$; CB-ELSA(black squares)\protect\cite{eta_pap}, TAPS\protect\cite{Krusche:nv},GRAAL\protect\cite{Renard:2000iv} and CLAS\protect\cite{Dugger:ft} data (in
light gray). The solid line represents the result of our fit.. 
For further details see~\protect\cite{eta_pap}.
 }
\vspace*{-0.2cm}
\label{fig_eta}
\end{figure}

\begin{figure}[t!]
  {\includegraphics[height=.205\textheight]{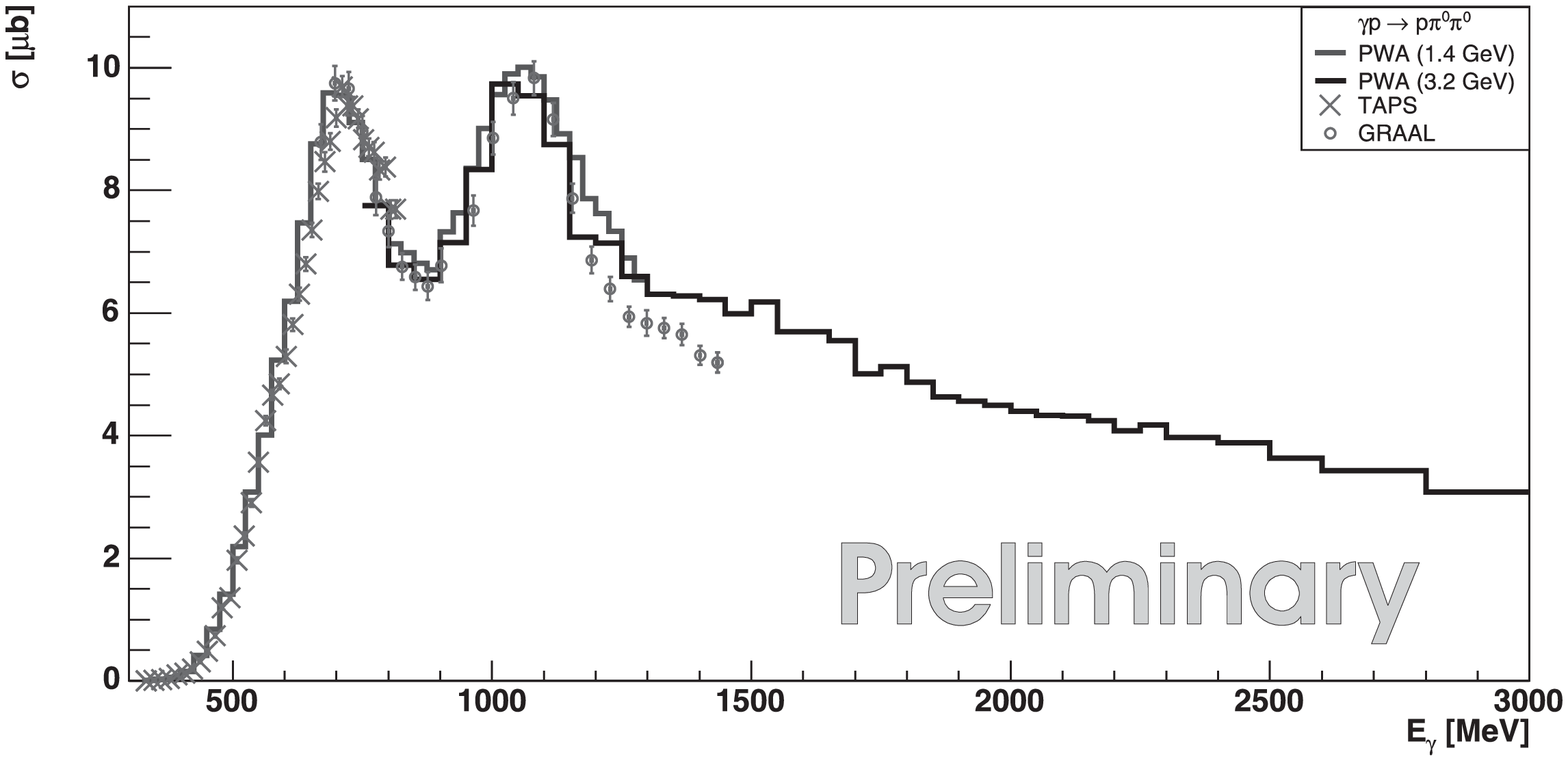}}
  {\includegraphics[height=.205\textheight]{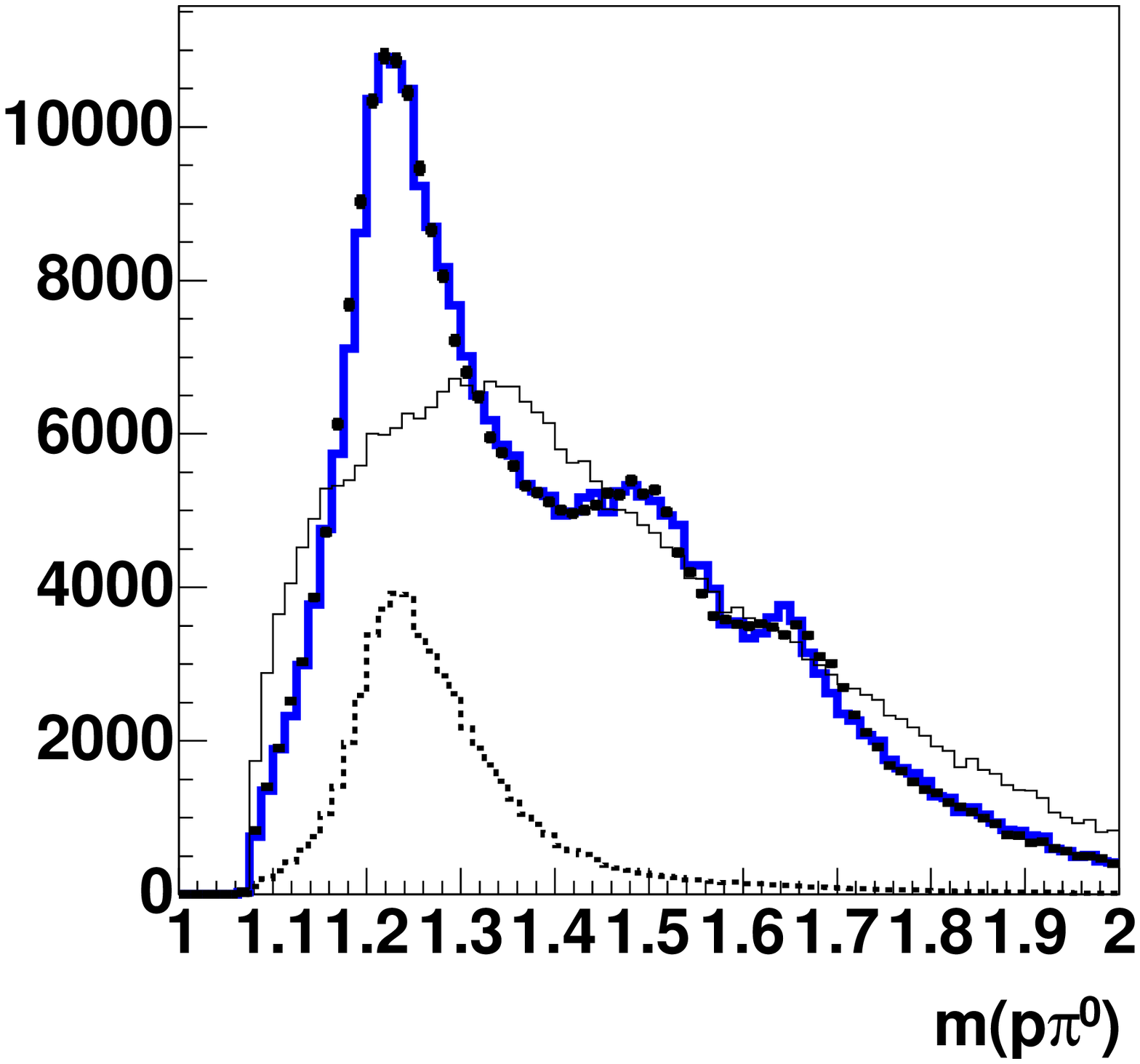}}
\caption{Left: Total cross section as obtained by integrating the result of
the PWA over phase space (solid line), in comparison
to preliminary TAPS\protect\cite{kottu} and GRAAL\protect\cite{graal}
data. Right: $p\pi^0$ invariant mass for $E_{\gamma}$=0.8-3.0~GeV  
in comparison to the result of the PWA. 
The plots shows the experimental data (points with error bars), the
result of the PWA  (solid gray curve), the contribution of the $\rm
  D_{13}$(1520) (dashed black curve) and the phase
  space distribution (thin black line), preliminary.}
\vspace*{-0.2cm}
\label{xsec_high}
\end{figure}

At medium energies, our present understanding of QCD is still very
limited. Here, in the energy regime of meson and baryon 
resonances 
perturbative methods can no longer be applied. 
One of the key issues 
is therefore to identify
the relevant degrees-of-freedom and the effective forces between them. 
A necessary step towards this aim is undoubtedly a 
precise knowledge of the experimental spectrum of baryon resonances
and of their properties.  
To search for new, or "missing" resonances is one of the aims of experiments
performed at ELSA.  
Experiments with electromagnetic probes are of course not only interesting to
search for unknown states but also to determine the properties of resonances
in general. 
The properties of a resonance are also of big importance for an
interpretation of its nature. One immediate debate in the light of the
possible observation of a pentaquark is e.g. whether the $\rm P_{11}$(1710) and
the $\rm P_{11}$(1440) might be pentaquarks rather than 3-quark states. A good
understanding of their production and decay properties may help to elucidate
their nature. 
At ELSA the photoproduction of different final states
has been investigated. To provide an overview, total cross sections
measured by the SAPHIR and CB-ELSA experiment are summarized in 
Fig.~\ref{all_wq}.  
\begin{figure}
\centerline{\psfig{file=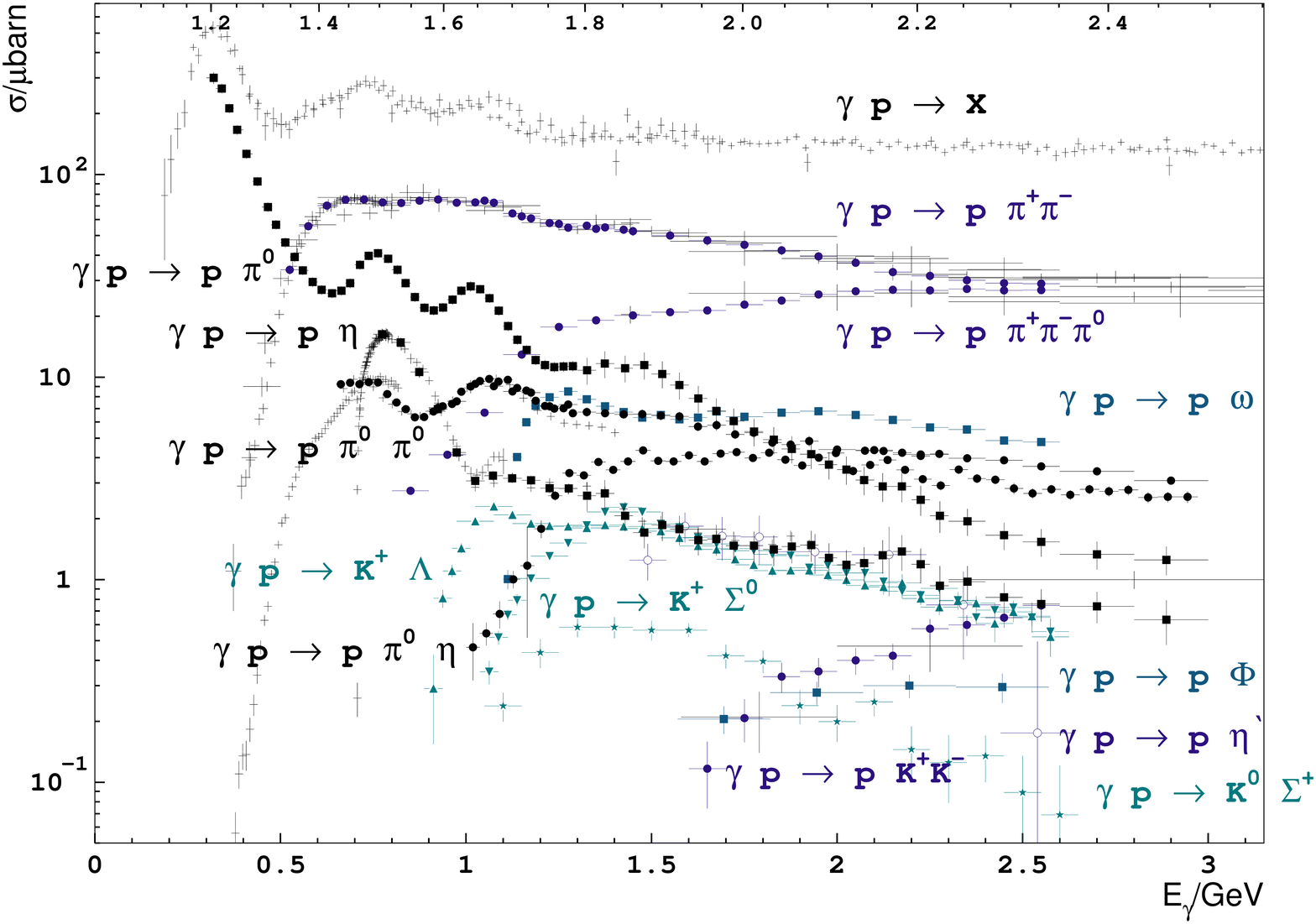,width=0.999\textwidth}}
\vspace*{-0.2cm}
\caption{Total cross sections for various final states measured at
  ELSA (filled and empty symbols) in comparison with the results from
  other
  experiments\protect\cite{pdg,graal,Krusche:nv,Renard:2000iv,Dugger:ft,kottu,wqs_others}
  (crosses). Data taken at ELSA: The 
   $\gamma p \to p \pi^0$, $\gamma p \to p \eta$, $\gamma p
  \to p \pi^0\pi^0$ and the $\gamma p \to p \pi^0\eta$ cross sections
  shown  have   been   measured by
  CB-ELSA\protect\cite{Bartholomy:04,eta_pap,cbelsa_ppi0pi0,igor}, the 
  other single and multi-meson cross sections by SAPHIR\protect\cite{saphir_all,saphir_old,klambda_saphir}.  
}
\vspace*{-0.2cm}
\label{all_wq}
\end{figure} \\[+3ex]
{\bf SAPHIR} \\
An interesting example for the results obtained by SAPHIR, apart from the
possible observation of a pentaquark state\cite{theta_saphir}, is the 
$\gamma p \to K^+ \Lambda$-channel. The total cross section
shows two bumps at W=1700~MeV, and W=1900~MeV\cite{saphir_old}. 
The enhancement around 1900~MeV, which was observed by SAPHIR for the first time, was
interpreted by Mart and Bennhold as being due to a new
resonance\cite{mart_bennhold}. It was then identified
with a $\rm D_{13}$(1895), a state which would be in nice agreement with 
quarkmodel predictions.  
But its existence is still controversially discussed; in different models the
observed structure is explained by 
different processes\cite{klambda_interpret}.
Recently new high statistics data on this final state became available. 
SAPHIR\cite{klambda_saphir} and also CLAS\cite{klambda_clas} did provide new
data on cross sections and on the $\Lambda$ recoil polarization and  
LEPS\cite{klambda_leps} on the beam-asymmetry. 
The new data shows again an enhancement around 1900~MeV.
First prelimary results on an interpretation of the higher statistics
data have been shown by Mart and Bennhold\cite{mart_bennhold_conf}. 
Fitting the new SAPHIR data together with the beam asymmetry data from
LEPS they find that more than one resonance is needed to
describe the mass region around 1900~MeV. This work is still in progress, so
no conclusions on the existence of new resonances in the data can be drawn yet. \\[+3ex]
{\bf CB-ELSA } \\[+0.5ex]
{\bf The \boldmath$\gamma p \to p \eta$\unboldmath-channel} \\[+1ex]
Recently new data on $\eta$-photoproduction has been taken by the
CB-ELSA experiment, which extends the covered angular
and energy range significantly compared to previous measurements\cite{eta_pap}.
The $\eta$ was observed either in its $\gamma\gamma$- or 3$\pi^0$-
decay. 
The invariant masses show a clear $\eta$ signal over
an almost negligible background (Fig.~\ref{fig_eta}). 
The total cross section is also shown in Fig.~\ref{fig_eta} in
comparison to the TAPS\cite{Krusche:nv}, GRAAL\cite{Renard:2000iv}, and
CLAS\cite{Dugger:ft} data.  It was obtained by integrating the differential
cross sections using the result of the partial wave analysis (PWA) discussed
below, as an extrapolation to forward and backward angles.  
The PWA is necessary to extract the contributing resonances 
from the data. Its result is shown as   solid line  in
Fig.~\ref{fig_eta}. 
In the fit the following data sets were included in addition to the CB-ELSA
data on $\rm \gamma p \rightarrow p\eta $: The CB-ELSA data on $\rm \gamma p
\rightarrow 
p\pi^0 $ \cite{Bartholomy:04}, the TAPS data on $\rm \gamma p
\rightarrow p\eta $ \cite{Krusche:nv}, the beam asymmetries
$\rm\Sigma(\gamma p \rightarrow p\pi^0)$ and $\rm\Sigma(\gamma p
 \rightarrow p\eta)$ from GRAAL \cite{GRAAL2}, 
and $\rm\Sigma(\gamma p \rightarrow p\pi^0)$ and 
$\rm \gamma p \rightarrow n\pi^+ $ from SAID. 
Apart from known resonances a new state was found, a $\rm
D_{15}(2070)$ 
with a mass of ($2068\pm 22$)\,MeV and a width of ($295\pm 40$)\,MeV. Its
rather strong contribution to the data set is also shown in
Fig.\ref{fig_eta}. In addition an indication for a possible new $\rm
P_{13}(2200)$ was found. 
No evidence was found for a third S$_{11}$ for which claims have been
reported at masses of 1780\,MeV\cite{Saghai:2003ch} and
1846\,MeV\cite{Chen:2002mn}. \\[+3ex]
{\bf The \boldmath$\gamma p \to p \pi^0\pi^0$\unboldmath-channel }\\[+0.5ex]
The $\gamma p \to p \pi^0\pi^0$ cross section was measured by TAPS\cite{taps}
in the low energy range and by GRAAL\cite{graal} up to an incoming photon
energy of about 1500~MeV; two peak-like structures are observed. 
The data has been interpreted within the Laget-\cite{laget_graal} and Valencia
model\cite{oset_graal_pap}, 
resulting in very different interpretations. In the Valencia-model, which is
limited to the low 
energy region, the $\rm D_{13}$(1520) decaying into $\Delta (1232)\pi$
dominates the lower energy peak, while
in the Laget-model the  $\rm P_{11}$(1440) decaying into $\sigma p$ is clearly
the dominant contribution. \\
Recently data on $\rm \gamma p\to p \pi^0\pi^0$ has also been taken by the
CB-ELSA experiment extending the covered energy range up
the $E_{\gamma}$=3.0$\,$GeV\cite{cbelsa_ppi0pi0}. 
A PWA has been done to extract the contributing resonances and their 
properties from the data. The formalism used is summarized in\cite{formalism}. 
The fit uses Breit-Wigner 
resonances and includes  $s$- and $t$-channel amplitudes. 
An unbinned maximum-likelihood fit was performed which 
takes all the 
correlations between the five independent variables correctly into account. 
The fits include the preliminary TAPS data\cite{kottu} in the low energy region
in addition to the CB-ELSA data. 
Resonances with different quantum numbers were
introduced in various combinations allowing, so far, for the following decay
modes: $\Delta(1232)\pi$, $\rm N(\pi\pi)_s$, $\rm P_{11}(1440)\pi$, $\rm
D_{13}(1520)\pi$ and $\rm X(1660)\pi$.
For a good description of the data resonances like e.g. the $\rm
P_{11}(1440)$, the $\rm D_{13}(1520)$, the $\rm D_{13}/D_{33}(1700)$, the $\rm
P_{13}(1720)$, the $\rm F_{15}$(1680) as well as several additional states are
needed. 
One preliminary result of the PWA is a dominant
contribution of the  $\rm D_{13}(1520)\to \Delta \pi$
amplitude in the energy range, where the first peak in the cross
section occurs. 
Fig.~\ref{xsec_high} shows the total cross section obtained by fitting
the CB-ELSA and the TAPS data and by integrating the result of
the combined fit over phase space.  
In the CB-ELSA data baryon resonances not only decaying into $\Delta \pi$ but
also via $\rm D_{13}(1520)\pi$ and $X(1660)\pi$ are observed for the first
time.  
The enhancements at the corresponding $p\pi$ invariant masses
are clearly visible in Fig.~\ref{xsec_high}. The observation of baryon
cascades is 
also interesting with respect to 
the search for states which might not couple to $\pi N$ and $\gamma p$; they
still could be produced in such baryon cascades. \\[+3ex]  
{\bf CB-ELSA/TAPS} \\
For the data taking period from 9/2002 until 12/2003, 90$\rm CsI$
crystals have been removed in forward direction to open up a forward region
$\pm 30^{\circ}$ which was then covered by the TAPS detector consisting out
of 528~$\rm BaF_2$ crystals. 
With this setup, data with unpolarized and linearly polarized photons
has been
taken using a liquid hydrogen target. The analysis of the 
data is in progress. \\
In addition also data with solid targets has been taken to investigate 
possible medium modifications of mesons in the nuclear medium. 
The question of interest is here the origin of mass. 
Hadrons acquire mass because of chiral symmetry breaking. Due to the
symmetry  breaking the vacuum gains a complex structure  
($<q\bar{q}> \ne 0$) and the hadrons gain mass by interacting with this
vacuum. The expectation value of the quark-condensate is now expected to
decrease with increasing density and/or temperature. Therefore one
would expect a partial restoration of chiral symmetry in the nuclear medium
resulting also in a change of the meson masses. 
First preliminary results on $\gamma Nb  \to \omega X, \omega \to \pi^0\gamma$
indeed indicate an enhancement at lower $\omega$-masses\cite{trnka}. The
found shape of the signal is in good agreement with 
calculations\cite{muehlich} assuming an in-medium modification of the $\omega$.  

\section{Future}
After the TAPS detector has left Bonn it is of course necessary to close the
forward angular range of $\pm 30^{\circ}$ again. 
This will be achieved by a
forward detector plug consisting out of the 90 $\rm CsI$-crystals used in the
original detector setup. Differently than the rest of the crystals, which are
read out by photodiodes the forward plug crystals will be read out by
photomultipliers. This new readout will allow to include the forward plug in
the first level trigger. In front of the crystals scintillator plates will be
placed to detect charged particles. 
In addition a Mini-TAPS array will cover the  forward angular range further down to 
$1.5^{\circ}$. Also here plastic scintillators are placed in front of the
crystals to allow the discrimination between charged particles and photons. 
With this detector setup single and double polarization experiments will be performed in
2005. These will help to discriminate between ambiguous solutions
in the PWA and will also provide a higher sensitivity to smaller
contributions. 
\\ 
For the further future it is planned to extent the trigger capabilities of 
the crystal barrel further to backward angles. This will allow to trigger with
high efficiency on rare channels and  on channels without any
charged particles in the final state measured. In addition it is also planned to
install a forward tracking system to be able 
to detect charged particles such as kaons in forward direction.

\section*{Acknowledgments}
The author acknowledges an Emmy Noether grant from the Deutsche
Forschungsgemeinschaft.

\end{document}